\documentclass{article}

     \usepackage[preprint]{neurips_2019}

\PassOptionsToPackage{numbers, compress}{natbib}

\usepackage[toc,page]{appendix}

\usepackage{rotating}

\usepackage{amsmath,amsfonts,bm}

\def\eqref#1{equation~\ref{#1}}

\def\1{\bm{1}}

\DeclareMathAlphabet{\mathsfit}{\encodingdefault}{\sfdefault}{m}{sl}
\SetMathAlphabet{\mathsfit}{bold}{\encodingdefault}{\sfdefault}{bx}{n}

\newcommand{\E}{\mathbb{E}}

\usepackage[utf8]{inputenc} 
\usepackage[T1]{fontenc}    
\usepackage{hyperref}       
\usepackage{url}            
\usepackage{booktabs}       
\usepackage{amsfonts}       
\usepackage{nicefrac}       
\usepackage{microtype}      
\usepackage{amsmath}

\title{DECoVaC: Design of Experiments with Controlled Variability Components}

\author{%
  Thomas Boquet\thanks{indicates equal contribution} \\
  Element AI \\
  \texttt{thomas@elementai.com} \\
   \And
   Laure Delisle\footnotemark[1] \\
   Element AI \\
   \texttt{laure.delisle@elementai.com } \\
   \AND
   Denis Kochetkov\footnotemark[1] \\
   Element AI \\
   \texttt{denis.kochetkov@elementai.com } \\
   \And
   Nathan Schucher \\
   Element AI \\
   \texttt{nathan@elementai.com} \\
   \And
   Parmida Atighehchian \\
   Element AI \\
   \texttt{parmida@elementai.com} \\
   \And
   Boris Oreshkin \\
   Element AI \\
   \texttt{boris@elementai.com} \\
   \And
   Julien Cornebise \\
   Element AI \\
   \texttt{julien@elementai.com} \\
}
\date{March 2019}

\begin{document}

\maketitle





\begin{abstract}
Reproducible research in Machine Learning has seen a salutary  abundance  of progress lately: workflows, transparency, and statistical analysis of validation and test performance. We build on these efforts and take them further. We offer a principled experimental design methodology, based on linear mixed models, to study and separate the effects of multiple factors of variation in machine learning experiments. This approach allows to account for the effects of architecture, optimizer, hyper-parameters, intentional randomization, as well as unintended lack of determinism across reruns.  We illustrate that methodology by analyzing Matching Networks, Prototypical Networks and TADAM on the miniImagenet dataset.

\end{abstract}

\section{Introduction}
\label{intro}

Concern about reproducible science has grown in the machine learning field in the past decade, with multiple studies finding that a significant proportion of published research could not be reproduced~\citep{Henderson2018DeepRL, Melis2018OnTS}.
To address this, the community has already come up with several recommendations for producing reproducible research along three main axes:
\begin{itemize}
  \item \textbf{Workflows:} In their work, \cite{chen2018open, Cebrat2018BuildingAR, Tatman2018APT} describe workflows and processes for consuming results from other labs, as well as initial data capture processes required for reproduction.
  \item \textbf{Transparency efforts:} Vowing for transparency, \cite{chen2018open, Henderson2018DeepRL, Cebrat2018BuildingAR} advocate for the release of the models' hyper-parameters and the method by which they were selected.
  \item \textbf{Statistical tools:} Ablation studies \citep{lipton2018troubling}, significance testing and error analysis \citep{Reimers2017ReportingSD, Henderson2018DeepRL, Melis2018OnTS}, measure of performance and visual inspection over factors of variation (hyper-parameters, regularization, random seed, optimization regime) using bootstrap confidence bounds and power analysis~\citep{Henderson2018DeepRL, Melis2018OnTS, lipton2018troubling}.
\end{itemize}

We believe that these salutary initiatives, focused on studying validation and test performance through observation only, can and should be systematized and taken further.

In this work, we offer a principled methodology for evaluating the variability of machine learning experiments. More specifically, we propose an experimental design methodology for statistical tests of variability in model performance across runs and factors. This systematic statistical analysis addresses the failure to identify the factors of empirical variation highlighted by \citet[Section 3.2]{lipton2018troubling}. Done across- and within-models, we argue that it benefits hyper-parameters selection, clarifies the stability of architectures, and strengthens state-of-the-art claims.

To illustrate our methodology, we then apply it to a suite of prominent algorithms in the domain of few-shot learning, a key area for applications of machine learning in data-poor environments~\citep{feifei2006oneshot}. Low data availability and quality arguably put results at a higher risk of uncontrollable variability, making  few-shot learning algorithms a fit example for our methodology.

\section{Methodology}
\label{Methodology}

\subsection{Experimental design}
Our experimental design methodology consists of three steps: establishing research hypotheses, gathering data using a random re-seeding framework, testing the research hypotheses by fitting a linear mixed model on that data. 

\textbf{Hypotheses} The first step of the procedure we follow is to propose a set of research hypotheses regarding the stability of the algorithm across different experimental conditions. Each null hypothesis to be tested should assume stability, which can then be rejected if variations are statistically significant.

\begin{itemize}
    \item H1: The results are stable across all runs of the same model and same optimizer using the same hyper-parameters configurations but the same random seeds.
    \item H2: The results are stable across all runs of the same model and same optimizer using the same random seeds but using different hyper-parameters configurations.
    \item H3: The results are stable across all runs of the same model and same optimizer using the same hyper-parameters configurations and different random seeds.
\end{itemize}

\textbf{Re-seeding framework} The second step is to design experiments to generate the data and test our hypotheses. We sample hyper-parameters configurations at random for each given implementation of each algorithm. We set the seeds of the pseudo-random number generators (PNRG) to ensure similar streams for multiple runs\footnote{In an ideal world we would use PRNGs designed for parallel pseudo-independent streams accross experiments. However for practical purposes, as such PRNGs are rarely available in common ML libraries, we deemed acceptable to simply use multiple seeds of quality PRNGs such as Mersenne Twister \citep{matsumoto1998mersenne}.}. 

\begin{figure}[htbp]
\includegraphics[width=.7\textwidth]{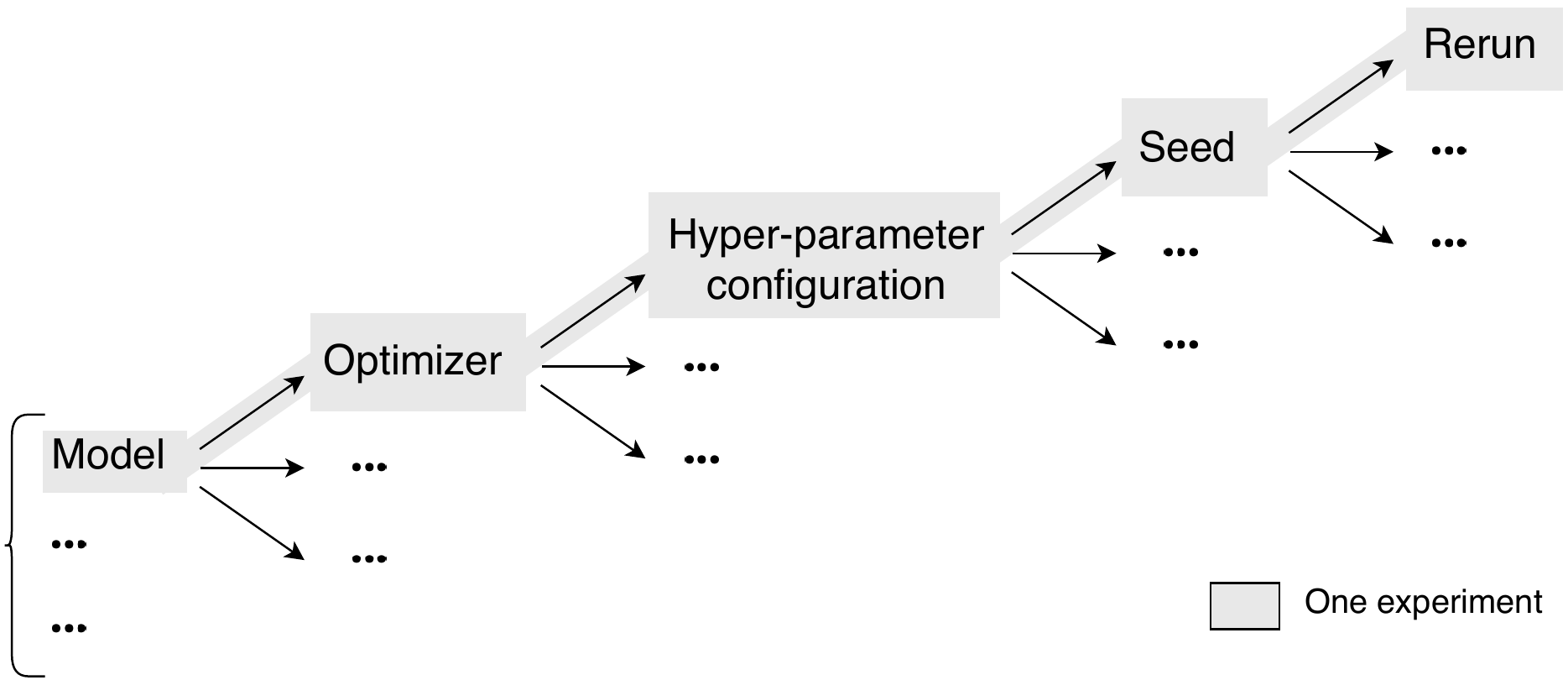}
\centering
\caption{Visualization of tree of experiments and PRNG seeding procedure.}
\label{fig:tree}
\end{figure}

As illustrated in Figure \ref{fig:tree}, for each combination of model and optimizer we sample a fixed number of hyper-parameter configurations. For each configuration, we s and rerun the training a fixed number of times for each configuration. An experiment consists of a rerun for a given combination of model, optimizer, PRNG seed, and sampled hyper-parameter configuration. We use the term \textit{rerun} to indicate the act of running an algorithm with exactly the same hyper-parameter settings and the same random seed multiple times in the context of an experiment.

\textbf{Statistical tests} The third step of the procedure is to select and use statistical tools to test our set of hypotheses. To do so we define a statistical model based on a linear mixed model suited to analyze clustered data. We then fit this model on the data and perform statistical tests on that model.

\subsection{Linear Mixed Models}

To measure the difference of means between groups, one can use linear models \citep{fisher1919xv, bates2004linear} or Bayesian linear models \citep{gelman2006data}. Linear models is a wide family of statistical models to measure the impact of different factors on a target variable. Two sample t-tests are a simple form of Analysis of variance (ANOVA) with one binary factor. It can also be seen as linear regression with a binary explanatory variable and no bias. To compare more than two groups w.r.t the same target variable, it's possible to use a one-hot encoded categorical variable. In our setting, we can assume we are in presence of noisy clustered observations of the target variable. 

Since we control almost completely the environment where the experiments are run and thus the data-generating process, we can define a specific design to reason about statistical reproducibility while comparing the results of different runs of different algorithms.
For each sample, we retrieve the information about the experiment name, the hyper-parameter configuration, the random seed used, the rerun identifier and the test accuracy on the meta-test split.
 
\begin{figure}[t]
\includegraphics[width=1.\textwidth]{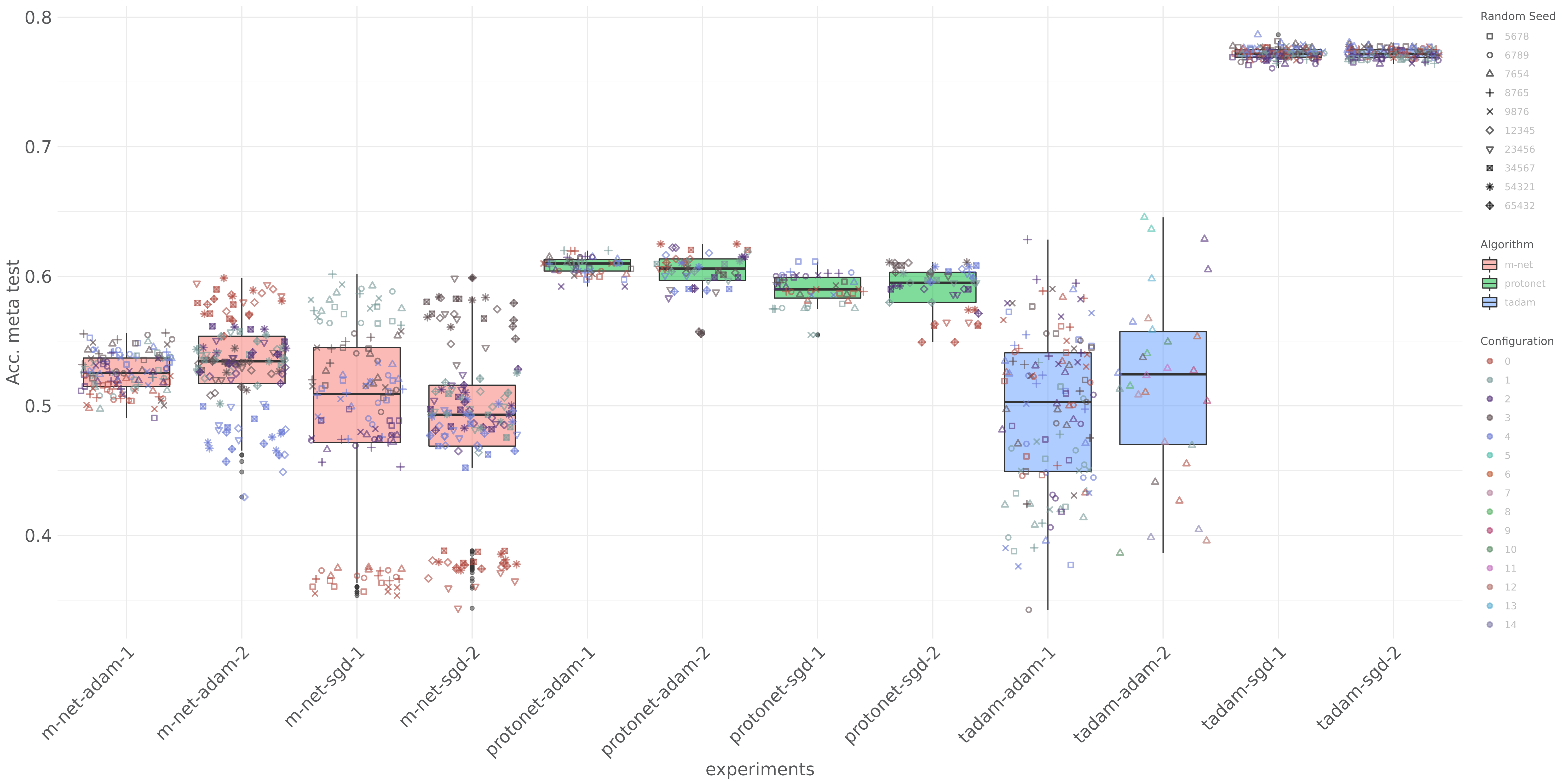}
\centering
\caption{Distribution of results of reruns over models, optimizers, hyper-parameter, and PRNG seed. The color of the boxplots represents the model, the color of the markers the hyper-parameters configuration and the different markers the random seed used in each trial. We illustrate each combination of model/optimizer/hyper-parameters for two sub-sets of seeds.}
\label{fig:ditribution_results}
\end{figure}

In our setup, we have $N\times D$ features $\mathbf{X}$ corresponding to a contrast matrix (one-hot encoded vector for the couple model-optimizer). $N$ is the total number of experiments, i.e. of leaves in the tree of Figure~\ref{fig:tree}. $D$ is the number of distinct experiments illustrated in Figure~\ref{fig:ditribution_results}, i.e. the number of combinations of models and optimizers multiplied by the number of reruns per seed. In other words, $D$ is the number of leaves in the tree integrated over hyper-parameters and seeds.

In the simple case where we assumed that our observations are i.i.d, we could estimate the effects of each experiment with the linear regression model\footnote{Note that this linear model is equivalent to a one way ANOVA with i.i.d samples.}:

\begin{equation}
\mathbf{y} = \mathbf{X}\boldsymbol{\beta} + \alpha + \epsilon,
\end{equation}

where $\boldsymbol{\beta} \in \mathbb{R}^D$ is the slope vector, $\alpha\in\mathbb{R}$ is the intercept, and $\epsilon\sim\mathcal{N}(\mathbf{0}, \mathbf{I})$ is random noise. In our setup, $\boldsymbol{\beta}$ and $\alpha$ are "fixed effects": we want to measure the difference between groups with constant effects across our dataset $(\mathbf{X}, \mathbf{y})$. To achieve this, we can maximize the likelihood $\mathbf{y} \sim \mathcal{N}(\mathbf{X}\boldsymbol{\beta} + \alpha, \mathbf{I})$ to find point estimates of $\boldsymbol{\beta}$ and $\alpha$ that fit the data.

However, with our design, we know that there is a structure in the data generating process and that the data $(\mathbf{X}, \mathbf{y})$ is therefore not i.i.d. To circumvent this modeling problem, we can rewrite our linear model by introducing normally-distributed ``random effects'' $ \mathbf{b}$ that vary across the population:
\begin{align}\label{eq:1}
 \mathbf{b} &\sim \mathcal{N}(\mathbf{0}, \sigma^2 \mathbf{I}) \\
 \label{eq:2}
\mathbf{y} &= \mathbf{X}\boldsymbol{\beta} + \mathbf{Z}\mathbf{b} + \alpha + \epsilon,
\end{align}
The term $\mathbf{Z}\mathbf{b}$ models the clusters, where $\mathbf{Z}$ is the $N \times Q$ model matrix for the $Q$-dimensional random-effects variable $\mathbf{b}$.  In this setting we can rewrite the \textit{conditional} distribution:
\begin{align}
    \mathbf{y}|\mathbf{b} \sim \mathcal{N}(\mathbf{X}\boldsymbol{\beta} + \alpha + \mathbf{Z}\mathbf{b}, \sigma^2\mathbf{W^{-1}}).
\end{align}

Since $\E[\mathbf{b}] = 0$,  the dependent variable mean is captured by $\mathbf{X}\boldsymbol{\beta} + \alpha$ when we marginalize over all the samples. The random effects component $\mathbf{Z}\mathbf{b}$ captures variations in the data. It can be interpreted as an individual deviation from the group-level fixed effect.

In our context, we can write the model as follows:
\begin{align}
\mathbf{y}_{ijk} &= \boldsymbol{\beta} \mathbf{X}_{i} + \alpha +  \varepsilon_{ijk} \label{eq:3} \\
\varepsilon_{ijk} &= \mathbf{b}_{0j} + \mathbf{b}_{1k} + \epsilon_{i} \label{eq:4} \\
\mathbf{b}_{uj} & \sim \mathcal{N}(0, \sigma_{uj}) ,\qquad u\in\{1,2\} \label{eq:5} \\
\epsilon_{i} & \sim \mathcal{N}(0, \sigma_{\epsilon}) \label{eq:6}
\end{align}
where $A$ is the intercept representing the mean of the whole group, $\boldsymbol{\beta}$ is a vector of parameters representing the deviation of each experiment from the mean, and $Experiment_{i}$ is a one hot vector of experiments for the observation $i$. In this model, we can decompose the error into multiple structured components. To this end we compose it in three terms: $\mathbf{b}_{0j}$ is a random effect associated with an observation from a random seed $j$, $\mathbf{b}_{1k}$ is associated with an observation from a hyper-parameters configuration $k$ and $\epsilon_{i}$ is Gaussian noise. It is possible to regroup all the random intercepts as nuisance parameters in $\varepsilon_{ijk} = \mathbf{b}_{0j} + \mathbf{b}_{1k} + \epsilon_{i}$.

\subsection{Hypotheses testing}
\label{subsec:hypotest}
From the configurations used in the random re-seeding framework, we fit the linear mixed model defined in \eqref{eq:5}. As defined in our three hypotheses, our goal is to quantify the variability in the error $\alpha_{0j}$ linked to the seeds, quantify the variability in the error $\alpha_{1k}$ linked to the hyper-parameters configuration and estimate the differences in performances $\beta$ between the different experiments and algorithms.

We first perform likelihood ratio tests for each random effect added to the model to test H1 and H2, i.e. if adding any of the random effects significantly changes the likelihood of the model given the data.  
Rejecting H1 and H2 would indicate that the implementations' performances vary significantly with a change of seed or hyper-parameters configuration.

To address H3, we first need to test if a difference exists between all the reruns' mean for a given combination of model and optimizer. We can use an ANOVA with a correction for the degrees of freedom for the number of comparisons performed \citep{kenward1997small}. We finally compare the means of reruns of the same combination of model and optimizer, by computing the means' difference and providing standard errors and 95\% estimated confidence intervals of our estimators.

To estimate the parameters of the linear mixed model defined in \eqref{eq:5}, we use the R implementation provided by the \texttt{lme4} package \citep{bates2014fitting}. The estimates for the random effects and the fixed effects estimated with \texttt{lme4} can be augmented with the \texttt{lmerTest} package \citep{kuznetsova2017lmertest} to add corrected degrees of freedom for the p-values \citep{kenward1997small, satterthwaite1946approximate} in small samples settings to compare several groups.

\section{Experiments}
\label{Experiments}

We selected three prominent articles which present three key few-shot learning algorithms using metric-based learning. We study and compare their performance on a specific task and a specific dataset, common to all three algorithms. We investigate how they exhibit different behaviors subject to different random seeds reruns and hyper-parameter changes.

\subsection{Experimental protocol}
To illustrate our methodology, we follow the steps of experimental design described in Section \ref{Methodology}.

\textbf{Dataset}
We use the miniImagenet dataset proposed by \cite{vinyals16} to perform our experiments. To construct the tasks, we sample 5 classes uniformly and 5 training samples per class uniformly. We use the (meta-) train, validation and test splits from \cite{ravi2017iclr}.

\textbf{Models} For this review we have selected three metric-based few-shot learning models: Matching Networks~\citep{vinyals16}, Prototypical Networks~\citep{snell17}, and TADAM~\citep{boreshkin18}. These models represent the state of the art in the 5-shot case for 2016, 2017, and 2018, respectively. We identified the official or community-endorsed implementation for each model.

\textbf{Optimizers} 
We select two optimizers (Stochastic Gradient Descent, ADAM) that we use for every model. Using the same 10 PRNG seeds for each optimizer and model, we sample 15 hyper-parameters configurations using the distributions in Table \ref{table:hparams} for each seed. We rerun the training 10 times for each combination of model, optimizer, seed and hyper-parameters configuration. This amounts to 3,000 experiments per model. Overall, we evaluate 3 models for a total of 9,000 experiments.

\begin{table}
\centering
\caption{\small Search space for the experiments' hyper-parameters for Adam and SGD optimizers.}
    \centering
 \begin{tabular}{l c c c c}
            \toprule
            \textbf{Algorithms} & \textbf{TADAM} & \textbf{Proto nets} & \textbf{Matching nets} \\
            \midrule
            \textbf{Learning rate} & $\mathcal{U}$(0.1, 0.02) & $\mathcal{N}$(0.005, 0.0012) & $\log\mathcal{U}$(0.0001, 0.1) \\
            \textbf{LR decay rate} & $\mathcal{N}$(10, 1) & 0.5 & $\log\mathcal{U}$(0.00001, 0.01) \\
            \textbf{LR decay period (batch)} & 2500 & $\mathcal{U}$(500, 2000) & 1 \\
            \textbf{Query shots per class} & $\mathcal{U}$\{16, 64\} & 15 & $\mathcal{U}$(5, 30) \\
            \textbf{Pre-train batch size} & $\mathcal{U}$\{32, 64\} & - & - \\
            \midrule 
            \textbf{N-Way} & 5 & 5 & 5 \\
            \textbf{N-Shot / support set} & 5 & 5 & 5 \\
            \textbf{Number of tasks per batch} & 2 & 1 & 1 \\
            \textbf{Batch size} & 100 & 100 & 500 \\
            \textbf{Early stop (epochs)} & & - & 20 \\
            \textbf{Training steps (batches)} & 21K & 10K & 75K \\
            \textbf{Test episodes} & 500 & 600 & 600 \\
            \bottomrule
         \end{tabular}
    \label{table:hparams}
\end{table}

From those experiments, we fit the linear mixed model defined in \eqref{eq:5}. Our goal is to quantify the variability in the error $\mathbf{b}_{0j}$ linked to the seeds, quantify the variability in the error $\mathbf{b}_{1k}$ linked to the rerun and estimate the differences in performances $\beta$ between the different experiments and algorithms.

\subsection{Experimental results}
\label{ExperimentalResults}

Following our methodology described in Section \ref{subsec:hypotest}, we perform likelihood ratio tests for each random effect added to the model, see Table \ref{table:randomeffects}. We reject both H1 and H2 related respectively to the seed and the hyper-parameters configuration, and confirm that the implementations' performances vary significantly with those factors. Intuitively, it means there is enough correlation between the observations using the same random seed and the same configuration to estimate co-variance parameters, see Table \ref{table:randomeffectsparams}.

\begin{table}[h!]
\centering
\caption{\small Random effects ANOVA.}
    \centering
 \begin{tabular}{l c c c c c c}
            \toprule
\quad &	\textbf{npar} &	\textbf{logLik} &	\textbf{AIC} &	\textbf{LRT} &	\textbf{Df} &	\textbf{Pr(>Chisq)} \\
            \midrule
(1 | seeds) & 14 & 2493.680 & -4959.361 &	24.41725 & 1 &	7.757113e-07 \\
(1 | hparams) &	14 & 1854.749 &	-3681.498 &	1302.27988 & 1 &	3.612198e-285 \\
            \bottomrule
         \end{tabular}
    \label{table:randomeffects}
\end{table}

We then test if a difference exists between all the experiments, and find that there is significant difference in the experiments' accuracy means as presented in Table \ref{table:anovamixed}. We finally compare the means of reruns of given combinations of model and optimizer by computing the means' difference and providing standard errors and 95\% estimated confidence intervals of our estimators: see Table \ref{table:meansdiff}. Among all the comparisons, we do not find any statistical difference. In that sense we do not reject H3 and confirm that rerunning the same combination or model and optimizer using the same hyper-parameters configuration and same random seed yields stable results.

\subsection{Benefits}
Based on the hypotheses discussed in Section \ref{subsec:hypotest} , our experimental design not only suggest a specific setup for reproducible results but also identifies the factors of variation explicitly in case of failure in reproducing the same results by giving quantified measurement for the effect of variation on the results of the model. 

\begin{table}[h!]
\centering
\caption{\small Random effects parameters.}
    \centering
 \begin{tabular}{l c c c c c c}
            \toprule
\quad 	\textbf{Groups} &	\textbf{Name} &	\textbf{Variance} &	\textbf{Std. Dev.} \\
            \midrule
 seeds &   (Intercept) & 0.0000309 & 0.005559 \\
 hparams &  (Intercept) & 0.0017922 & 0.042334 \\
 Residual &         &    0.0004338 & 0.020828 \\
            \bottomrule
         \end{tabular}
    \label{table:randomeffectsparams}
\end{table}

\begin{table}[h!]
\centering
\caption{\small Linear Mixed Model fixed effects results.}
    \label{tab:agreement}
    \centering
 \begin{tabular}{l c c c c c c c}
 \toprule
\quad & \textbf{Sum Sq} &	\textbf{Mean Sq} &	\textbf{NumDF} &	\textbf{DenDF} &	\textbf{F value} &	\textbf{Pr(>F)} \\
\textbf{experiments} &	0.133368 &	0.0121244 & 11 &	56.75 &	27.95 &	5.896146e-19 \\
            \bottomrule
         \end{tabular}
    \label{table:anovamixed}
\end{table}

\begin{table}[h!]
\centering
\caption{\small Means comparisons, see Figure \ref{fig:ditribution_results} for graphical representation.}
    \centering
 \begin{tabular}{l c c c c c c c}
            \toprule
\quad & \textbf{Estimate} &	\textbf{Std. Error} &	\textbf{lower} &	\textbf{upper} &	\textbf{Pr(>|t|)} \\
m-net-adam & -0.006338 &	0.027132 &	-0.06071979 &	0.048044548 &	8.161823e-01 \\

m-net-sgd &	0.006836 &	0.0271328 &	-0.04754668 &	0.061219636 &	8.020148e-01 \\
protonet-adam &	0.004232 &	0.027331 &	-0.05051218 &	0.05897616 &	8.774987e-01 \\
protonet-sgd &	-0.000794 &	0.027342 &	-0.05555950 &	0.05397145 &	9.769353e-01 \\
tadam-adam &	-0.020189 &	0.02308498 &	-0.06631741 &	0.02594024 &	3.851393e-01 \\
tadam-sgd &	-0.000014 &	0.02713232 &	-0.05439609 &	0.05436825 &	9.995925e-01 \\
            \bottomrule
         \end{tabular}
    \label{table:meansdiff}
\end{table}




\section{Conclusion}

We model effects dependencies (model, optimizer, hyper-parameters configuration, seed, rerun) on the performance in a hierarchical manner: see \ref{fig:tree}. Exploiting the co-variance structure of the linear mixed models allows to establish relations between different factors like random seed, hyper-parameters reruns. By studying this, we directly test for significant differences in performance between various architecture choices subject to factors variation. This for instance allows for more informed decisions on architecture choices and training regimes, and gives better clarity on the variability characteristics of a given model.

To enable variability studies following our methodology, we intend to release a ready-to-run notebook (along with its docker container for easy reproducibility) with all the statistical tests implemented and an easy-to-follow example. This release will also include the data gathered in our case study to examplify the methodology.

\bibliography{repro}
\bibliographystyle{repro}

\newpage

\appendix

\end{document}